\documentclass[preprint,showpacs,preprintnumbers,amsmath,amssymb]{revtex4}

\usepackage{graphicx}
\usepackage{epstopdf}

\usepackage{dcolumn}
\usepackage{bm}
\usepackage{color}

\usepackage{feynmf}
\usepackage{slashed}
\usepackage{enumerate}
\usepackage{caption}

\usepackage[utf8]{inputenc}

\usepackage[colorlinks=true,urlcolor=blue,anchorcolor=blue,citecolor=blue,filecolor=blue,linkcolor=blue,menucolor=blue,linktocpage=true]{hyperref} 

\begin{document}

\title{Structures of the Massive Vector Boson Propagators at Finite Temperature Illuminated by the Goldstone Equivalence Gauge}



\author{Yi-Lei Tang}
\thanks{tangylei@mail.sysu.edu.cn}
\affiliation{School of Physics, Sun Yat-Sen University, Guangzhou 510275, China}
\affiliation{Quantum Universe Center, KIAS, 85 Hoegiro, Seoul 02455, Republic of Korea}


\date{\today}

\begin{abstract}
Inspired by the Goldstone equivalence gauge, we study the thermal corrections to an originally massive vector boson by checking the poles and branch cuts. We find that part of the Goldstone boson is spewed out from the longitudinal polarization, becoming a branch cut which can be approximated by the ``quasi-poles'' in the thermal environment. In this case, physical Goldstone boson somehow partly recovers. We also show the Feynmann rules for the ``external legs'' of these vector boson as well as the recovered Goldstone boson, expecting to simplify the vector boson participated process calculations by adopting the similar ``tree-level'' logic as in the zero temperature situation. Gauge boson mixing case are also discussed. Similar results are shown in other gauges, especially in the $R_\xi$ gauge.
\end{abstract}


\maketitle
\section{Introduction}

In the literature, it is well-known that the number of degrees of freedom of an originally massless photon or gluon in the thermal environment is different from those in the zero temperature. Due to the collective motion of the plasma particles, the longitudinal degree of freedom arises in the form of a quasi-particle. Such an oscillation mode is called a ``plasmon'' and has long been investigated (For the early works, see Ref.~\cite{PlasmonPioneer1, PlasmonPioneer2}. For some applications in the early universe, see Ref.~\cite{PlasmonApp1, LinTongYan}, and see Ref.~\cite{Fupo} for detecting the axion. Ref.~\cite{ThermalFieldBook} provided the systematic derivations). The structure of an originally massless vector boson at finite temperature can be studied through calculating the self-energy loop diagrams. In the thermal plasma, temperature-dependent mass corrections arise for both transverse and longitudinal propagators. The dispersion relations for the ``on-shell'' bosons in the plasma then become complicated, and are different between the transverse and longitudinal polarizations.

Transplanting these discussions directly to an originally massive vector boson faces difficulties. Decompose the propagator of the massive vector boson into transverse and longitudinal polarization contributions in a general $R_\xi$ gauge is formidable. Although the polarizations can be well-separated in the specific $\xi=0$ Landau gauge, however the relationship between the Goldstone boson and the longitudinal polarization are still puzzling, making it difficult to examine the ``on-shell'' properties of the vector bosons in the thermal plasma.

For practical calculations of physical processes, e.g., the dark matter annihilations into massive vector bosons, we can compute the imaginary part of the loop digrams to automatically sum over all the inclusive processes to elude the appearance of the vector boson's and Goldstone boson's external legs, or as were done in Ref.~\cite{LaineNeutrino3, BelowTcLaine2, BelowTcLaine3, BelowTcLaineNLO} for the calculations of the processes of the sterile neutrinos in the broken phase. These imaginary parts arise from the sum rules of the poles and branch cuts of the resummed internal propagators. A naively direct calculation with this method in the general $R_\xi$ gauge is plagued by the intricate tensor and analytical structures of the vector boson's propagators. The whole process seems to be a bunch of tedious integrations with the disconcerting ``physical meanings''. Motivated by this, we want to search for a reliable and straightforward tree-level method for a better ``physical picture'' to elude the complicated loop calculations. This is achieved by investigating the analytical structure, especially by enumerating and calculating the poles and branching cuts of the resummed massive vector boson's propagators, so this tree-level approach is mathematically equivalent to the lowest-order inclusive calculations depending on the arduous sum rules of all the pole and branching cut contributions. Our tree-level effort feliticiously classify these poles and branching cuts well and identify them to be the different polarizations' and Goldstone boson's contributions. Therefore it is more intuitive and simpler for one to begin with. Even if one might be obstinate to fix upon the traditional inclusive loop methods, he could still refer to our paper for a useful decomposition of the propagators and a classification of the poles and branching cuts correctly for a more concise and intruitive operation.	To anatomize the Goldstone boson and vector boson modes in detail, we begin with a physical gauge, by which the Goldstone boson contributions are well separated with the gauge part. It is then more convenient for us to observe how the vector boson ``eats'' the Goldsone degree of freedom \cite{PhysicalGauge1, ChenjunmouPhysicalGauge}. In the thermal environment, It is then beneficial for us to apply this gauge to look into the thermal effects on all of the modes respectively, since all these polarization mode contributions are well-separated.

In this paper, we start with the ``Goldstone equivalence gauge'' introduced by Ref.~\cite{GoldstoneEquivalenceGauge}. Although later the complete $R_\xi$ results are also displayed briefly. The detailed derivations which had been warped in our paper are actually more circuitous then and inspired by the similar processes of the Goldstone equivalence gauge. One of the most prominent achievement in this paper is to show how the longitudinal polarization of the vector boson will somehow gradually ``decouple'' with the Goldstone boson as we heat the system. Finally this mode becomes a pure ``plasmon'' as the massless vector boson in the high temperature limit. The once ``been eaten'' Goldstone boson can somehow resurrect inside the ``tachyonic branch cut'', and with the ``quasi-pole'' approximation suggested in this paper, we can regard this as a partial massless Goldstone boson and calculate their effects as the external lines.

\section{Lagrangian Adopted and the Zero Temperature Propagator Decompositions}

For simplicity, we rely on a $U(1)$ toy model with only one gauge boson $A^{\mu}$ and one complex Higgs boson $H$. Part of the Lagrangian is then given by
\begin{eqnarray}
\mathcal{L} \supset -\frac{1}{4} F_{\mu \nu} F^{\mu \nu} + D_{\mu} H^{\dagger} D^{\mu} H + V(H), \label{LOrigin}
\end{eqnarray}
where $F_{\mu \nu} = \partial_{\mu} A_{\nu} - \partial_{\nu} A_{\mu}$, $D_{\mu} = \partial_{\mu} - i g A_{\mu}$, with $g$ to be the gauge coupling constant, and $V(H)$ is the gauge-invariant potential of the scalar sector. We do not concern the details on $V(H)$, and only need to know that this induces a vacuum expectation value (vev) $v$ of the Higgs boson to break the gauge symmetry spontaneously. Therefore,
\begin{eqnarray}
H = \frac{v + h + i \phi}{\sqrt{2}},
\end{eqnarray}
where $h$ is the remained Higgs boson, and $\phi$ is the Goldstone boson. Then the Lagrangian becomes
\begin{eqnarray}
\mathcal{L} \supset -\frac{1}{2} \partial^{\mu} A^{\nu} \partial_{\mu} A_{\mu} + \frac{1}{2} \partial^{\mu} A_{\mu} \partial^{\nu} A_{\nu} + \frac{1}{2} m_A^2 A_{\mu} A^{\mu} - m_A A^{\mu} \partial_{\mu} \phi + \frac{1}{2} (\partial^{\mu} \phi)^2, \label{LBroken}
\end{eqnarray}
where $m_A = g v$. Besides the (\ref{LOrigin}), the vector boson might couple with other fields, which contribute to the thermal masses in the one-loop level. We just parametrize these contributions by the temperature dependent functions $\Pi_{L,T,S,U}(k)$ which will be defined in (\ref{SelfEnergyFormalism}). In the hard thermal loop (HTL) approximation, only the $\Pi_{T,L}$ dependent on a thermal mass parameter $m_E$ are significant. Since we do not study the details of the couplings, we neglect all of them in the Lagrangian.

Now we introduce the gauge $n_{\mu} A^{\mu} = 0$, where $n_{\mu} = (1, -\frac{\vec{k}}{|\vec{k}|})$ for the Goldstone equivalence gauge\cite{GoldstoneEquivalenceGauge} and $k$ is the four-momentum of a plain wave in the momentum space. It is easy to prove that
\begin{eqnarray}
n^{\mu} = \frac{\sqrt{k^2} \epsilon_{LU}^{\mu} (k) - k^{\mu}}{|\vec{k}|-k^0},
\end{eqnarray}
where $\epsilon_{LU \mu} (k) = (|\vec{k}|, k_0 \frac{\vec{k}}{|\vec{k}|})/\sqrt{k^2}$ is the usual longitudinal polarization vector in the unitary gauge. One can easily verify that $n^{\mu} n_{\mu}=0$, making it to be a kind of light-cone gauge to simplify the calculations.

We are then going to follow the detailed processes described in Ref.~\cite{ChenjunmouPhysicalGauge} to fix the gauge and decompose the propagators, and we also adopt all the conventions and contractions rules there for convenience. With the aid of the gauge-fixing term
\begin{eqnarray}
\mathcal{L}_{\text{gf}} = \frac{1}{2 \xi} 	(n^{\mu} n^{\nu} \partial_{\mu} A_{\nu})^2,
\end{eqnarray}
the complete form of the propagator within the Goldstone part becomes
\begin{eqnarray}
\langle (A^{\mu}, \phi), (A_{\mu}, \phi) \rangle = \frac{i}{k^2-m_A^2+i \epsilon} \left(
\begin{array}{cc}
-(g^{\mu \nu} - \frac{n^{\mu} k^{\nu} + k^{\mu} n^{\nu}}{n \cdot k} + n^2 \frac{k^{\mu} k^{\nu}}{(n \cdot k)^2} ) & i \frac{m_A}{n \cdot k} ( n^{\mu} - n^2 \frac{k^{\mu}}{n \cdot k} ) \\
-i \frac{m_A}{n \cdot k} (n^{\nu} - n^2 \frac{k^{\nu}}{n \cdot k} ) & 1-\frac{n^2 m_A^2}{(n \cdot k)^2}
\end{array} 
 \right) \label{FullPropagator}
\end{eqnarray}
in the $\xi \rightarrow 0$ limit. $n^2$ terms are retained to compare with the general form in Ref.~\cite{ChenjunmouPhysicalGauge}. The matrix is extended from 4-dimension to 5-dimension, with an extra Goldstone degree of freedom. We use $\mu \nu \dots$ to indicates the 4-dimensional indices, and $MN\dots=0,1,2,3,4$ to express the extended indices including the Goldstone degree of freedom. $M,N,\dots=4$ corresponds to the Goldstone component.

Now we define the transverse polarization vectors $\epsilon_s^{\mu}$ ($s=\pm$), which are exactly the same with the usual $R_\xi$ gauge ones. They satisfy
\begin{eqnarray}
\epsilon^0_{\pm} & = & 0, \nonumber \\
k \cdot \epsilon_{\pm} & = & 0, \nonumber \\
\epsilon_+ \cdot \epsilon_-^* &=& \epsilon_- \cdot \epsilon_+^* = 0, \nonumber \\
\epsilon_- \cdot \epsilon_-^* &=& \epsilon_+ \cdot \epsilon_+^* = -1.
\end{eqnarray}
For the special $k=(k_0, 0, 0, k_3)$ along the z-axis, $\epsilon_{\pm}(k) = \frac{1}{\sqrt{2}}(0, 1, \pm i,  0)$. The $\epsilon_{\pm}(k)$ of a general $k$ can be acquired by directly rotating from the z-direction case. Then transverse projection operator are defined by $P_T^{\mu \nu} = \sum\limits_{s = \pm} \epsilon_s^{\mu *} \epsilon_s^{\nu}$. It is easy to verify that
\begin{eqnarray}
P_T^{i j} &=& \delta_{i j} - \frac{k_i k_j}{|\vec{k}|^2}, \nonumber \\
P_T^{0 i} &=& P_T^{i 0} = P_T^{0 0} = 0, \label{ProjectorT}
\end{eqnarray}
where $i$, $j$ are the space coordinates.

Extend $\epsilon_s^{\mu}$ to $\epsilon_{s}^M = \left( \begin{array}{c}
\epsilon_{\pm}^{\mu} \\
0
\end{array} \right)$, and $P_T^{\mu \nu}$ to $P_T^{M N}$ where the extra elements are supplemented with zero. The factors in the matrix of (\ref{FullPropagator}) can be decomposed of
\begin{eqnarray}
& & \sum_{s= \pm} \epsilon_s^{M} \epsilon_s^{N *} + \left(
\begin{array}{cc}
\frac{k^2}{(n \cdot k)^2} n^{\mu} n^{\nu} & i \frac{m_{A}}{n \cdot k} n^{\mu} \\
-i \frac{m_{A}}{n \cdot k} n^{\nu} & 1
\end{array}
\right) \nonumber \\
&=& P_T + \left(
\begin{array}{cc}
\frac{k^2}{(n \cdot k)^2} n^{\mu} n^{\nu} & i \frac{m_{A}}{n \cdot k} n^{\mu} \\
-i \frac{m_{A}}{n \cdot k} n^{\nu} & 1
\end{array} \right). \label{DecomposePropagator}
\end{eqnarray}
Ref.~\cite{ChenjunmouPhysicalGauge} had illustrated that the second term in (\ref{DecomposePropagator}) can be decomposed into $\epsilon_L^M \epsilon_L^{N*}$ near the $k^2 = m_A^2$ pole, where $\epsilon_L^M = \left(\begin{array}{c}
-\frac{m_{A}}{n \cdot k} n^{\mu} \\
i
\end{array} \right)$. This is the longitudinal polarization vector in the Goldstone equivalence gauge. However, to find out the thermal mass corrections, we need to extend this decompositions to the off-shell bosons. Here we separate the second term in (\ref{DecomposePropagator}) into $P_L+P_G$, where the longitudinal projector operator $P_L$ and the Goldstone boson's projector operator $P_G$ are defined to be
\begin{eqnarray}
P_L &=&  \left(
\begin{array}{cc}
\frac{k^2}{(n \cdot k)^2} n^{\mu} n^{\nu} & i \frac{m_{A}}{n \cdot k} n^{\mu} \\
-i \frac{m_{A}}{n \cdot k} n^{\nu} & \frac{m_{A}^2}{k^2+i \epsilon}
\end{array}
\right), \label{ProjectorL} \\
P_G &=&  \left(
\begin{array}{cc}
0 &0 \\
0 & \frac{k^2-m_{A}^2+i \epsilon}{k^2+i \epsilon}
\end{array}
\right). \label{ProjectorG}
\end{eqnarray}
Different with Ref.~\cite{ChenjunmouPhysicalGauge}, we revise the definition of the ``longitudinal polarization vector'' to be
\begin{eqnarray}
\epsilon_L^M(k) = \left(\begin{array}{c}
-\frac{\sqrt{k^2}}{n \cdot k} n^{\mu} \\
i \frac{m_A}{\sqrt{k^2}}
\end{array} \right)	\label{LongitudinalDefinition}
\end{eqnarray}
for both on-shell and off-shell (at least for time-like) vector bosons, we can easily see if we neglect the $i \epsilon$ term,
\begin{eqnarray}
P_L^{M N} = \epsilon_L^M \epsilon_L^{N *}.	\label{PLDefinition}
\end{eqnarray}
Finally, the propagator can be decomposed to 
\begin{eqnarray}
\langle (A^{\mu}, \phi), (A^{\nu}, \phi) \rangle = \frac{i}{k^2-m_{A}^2+i \epsilon} (P_T + P_L + P_G),
\label{FullPropagatorProjector}
\end{eqnarray}
where the Goldstone boson's projector 's numerator $k^2 - m_A^2+i \epsilon$ will cancel the $k^2=m_{A}^2$ pole while contribute to another $k^2=0$ pole. This pole again cancels the $\frac{m_{A}^2}{k^2}$ element in the longitudinal polarization projector, leaving us no physical massless degree of freedom. Therefore we can see clearly how the Goldstone boson has been ``eaten'' by the longitudinal polarization of the vector boson.

\section{Thermal Effects Added} \label{ThermalAdded}

In the thermal environment, the propagator of any particle should be corrected by the distribution functions. Remember the vector bosons obey the Bose-Einstein distribution, so we define
\begin{eqnarray}
n_B (k_0) = \frac{1}{e^{\beta |k_0|} - 1},
\end{eqnarray}
where $\beta = \frac{1}{T}$ and $T$ is the temperature. The tree-level thermal propagator can be written in the ``diagonalized form'' (See Page 204 of Ref.~\cite{ThermalEarlyReview1} for the corresponding details)
\begin{eqnarray}
D_{a b}^{F, MN} (k) = U_{a c}(k) \left( \begin{array}{cc}
\frac{i}{k^2-m_A^2+ i \epsilon} & 0 \\
0 & -\frac{i}{k^2-m_A^2 - i \epsilon}
\end{array} \right)_{cd} U_{db}(k) (P_T + P_L + P_G)^{MN},
\end{eqnarray}
where $U$ is given by
\begin{eqnarray}
U(k) = \left( \begin{array}{cc}
\sqrt{1+n_B (k_0)} & \sqrt{n_B (k_0)} \\
\sqrt{n_B (k_0)} & \sqrt{1+n_B (k_0)} \end{array} \right).
\end{eqnarray}
The above propagator is calculated in the ``$\sigma = \beta /2$'' condition. This is convenient for computing the mass shift in the ``real-time formalism'', because the self-energy diagram can also be written in the ``diagonalized form''
\begin{eqnarray}
-i \Pi_{ab}^{MN} (k) = U_{a c}^{-1} (k) \left( \begin{array}{cc}
-i \overline{\Pi}^{MN} (k) & 0 \\
0 & (-i \overline{\Pi}^{MN} (k)^*)
\end{array} \right)_{cd} U_{db}^{-1}(k).
\end{eqnarray}
Therefore all of the $U(k)$ and $U^{-1}(k)$ cancels with each other inside the ``self-energy string'' diagrams, leaving only those in the beginning and the end. We will then decompose $\overline{\Pi}^{MN}(k)$ to see its temperature dependence.

\begin{figure}
\includegraphics[width=0.32\textwidth]{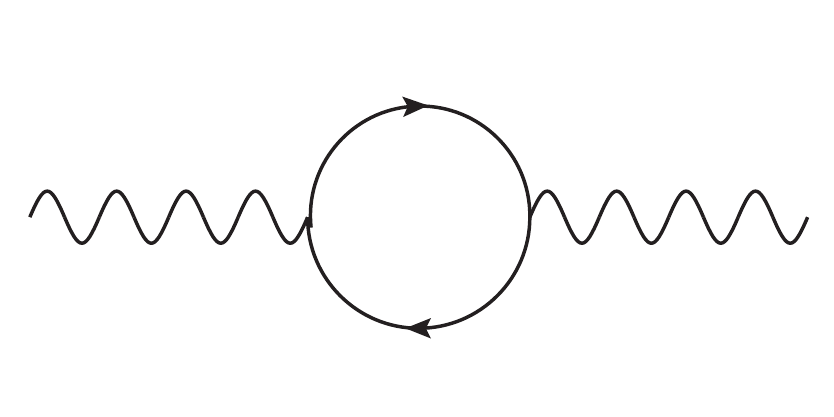}
\includegraphics[width=0.32\textwidth]{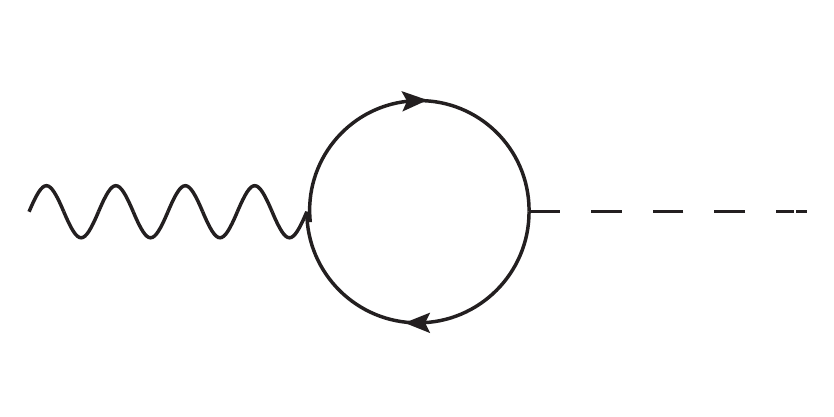}
\includegraphics[width=0.32\textwidth]{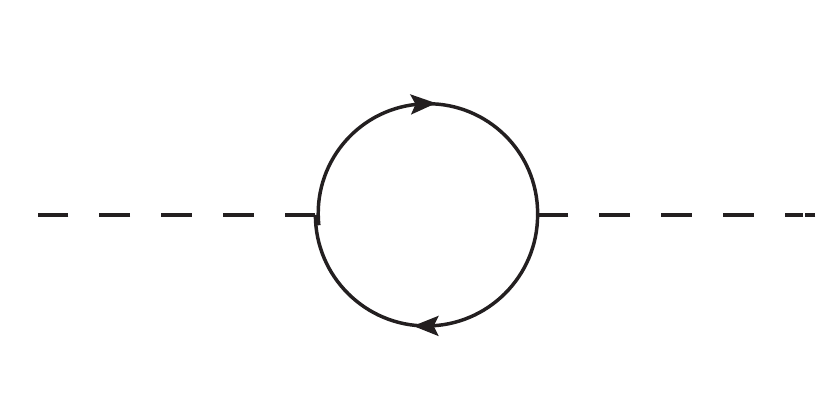}
\caption{Self-energy diagrams in $ZZ$, $ZG$, $GG$ propagators.}
\label{SelfEnergyDiagrams}
\end{figure}

\begin{figure}
\centering
\includegraphics[width=0.32\textwidth]{ZZPropagator.pdf}
\includegraphics[width=0.32\textwidth]{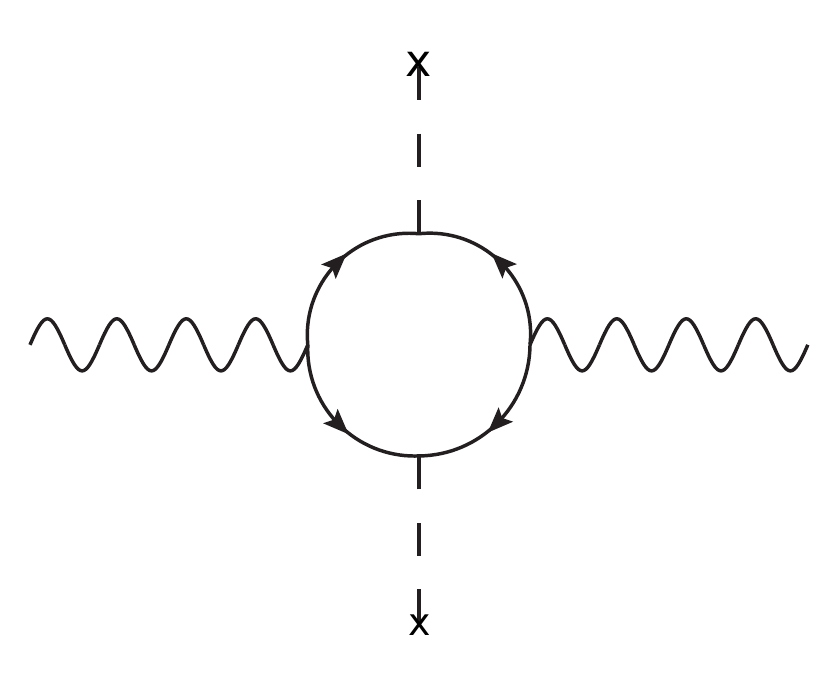}
\includegraphics[width=0.32\textwidth]{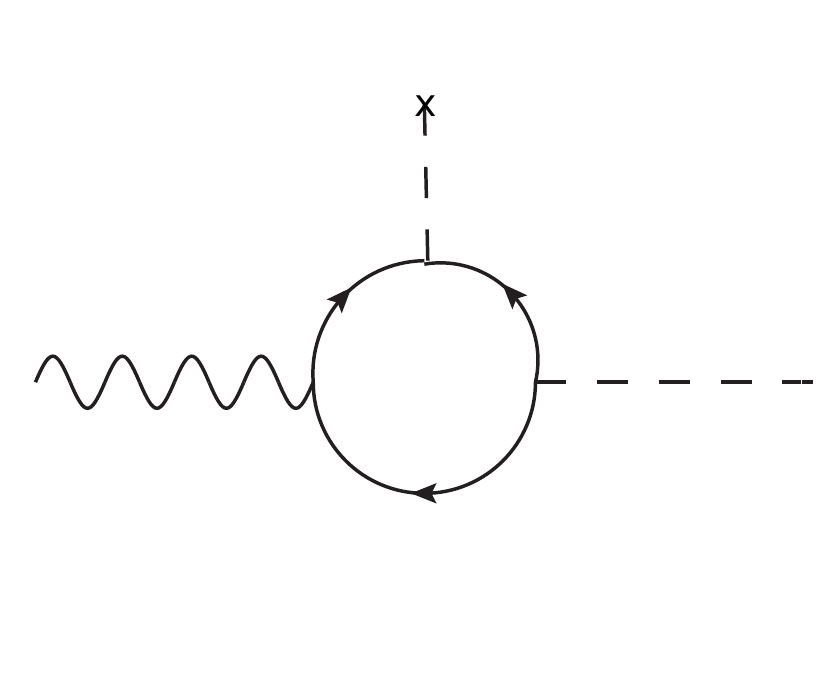}
\includegraphics[width=0.32\textwidth]{GGPropagator.pdf}
\includegraphics[width=0.32\textwidth]{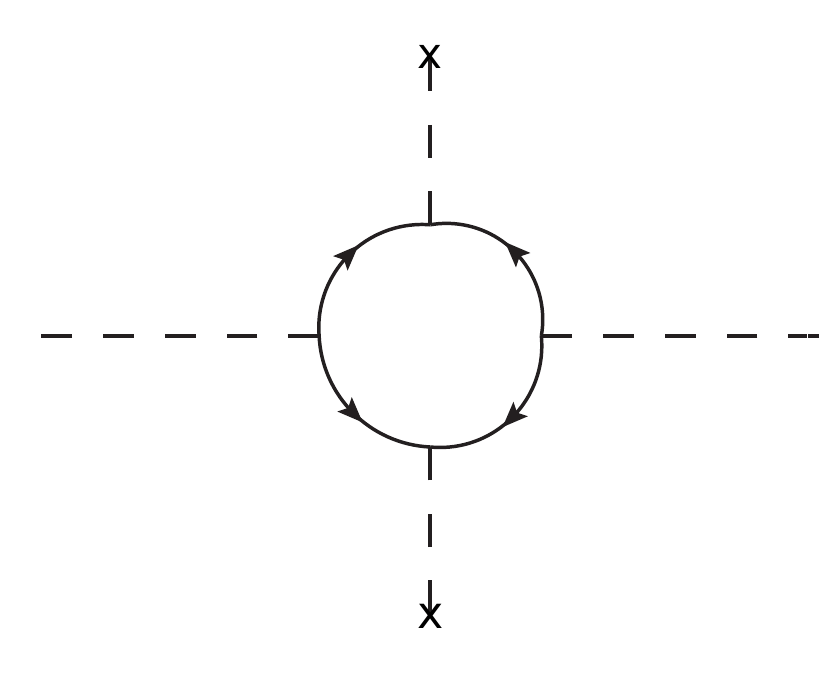}
\caption{Diagrams contributing to the self-energy diagrams before the $U(1)$ spontaneously breaking.} \label{RealDiagrams}
\end{figure}

In order to calculate the full thermal corrections on $\overline{\Pi}^{MN}(k)$, we need by principle to compute all the self-energy diagrams in Fig.~\ref{SelfEnergyDiagrams}.  At zero temperature, these three diagrams contribute to the $\delta m_1^2 A_{\mu} A^{\mu}$, $\frac{1}{2} \delta m_2 A^{\mu} \partial_{\mu} \phi$ and the $\delta m_3^2 \phi^2$ operators in the Lagrangian. Gauge symmetry requires $\delta m_1^2 = 2 m_A \delta m_2$ and $\delta m_3^2=0$ to preserve the renormalized tensor structures in  the (\ref{LBroken}) and (\ref{FullPropagator}). These relationships are guaranteed by gauge symmetry (or some formalism of Ward-Takahashi identity even in the broken phase, as will appear in (\ref{WTEquivalent}-\ref{CriticalEquation}).) and can be observed through the vev insertion diagrams in Fig.~\ref{SelfEnergyDiagrams}. The first diagram in Fig.~\ref{RealDiagrams} only contributes to the wave function renormalization of the gauge bosons without shifting its mass. The second and third diagram contribute to the $A_{\mu} A^{\mu} h^2$, and $A^{\mu} (\partial_{\mu} \phi) h$ respectively, therefore they work on $g^2$ and $g$ respectively. Finally, after inserting the vevs, $\delta m_1^2 = 2 m_A \delta m_2$ is equivalent to the $\delta g^2 = 2 g \delta g$. The last two diagrams in Fig.~\ref{RealDiagrams} corrects the potential $V(H)$, and these two diagrams do not disturb the gauge coupling constants. One aspect on these two diagrams, e.g., in the $V(H) = - m_h^2 H H^{\dagger} + \lambda (H H^{\dagger})^2$ situation, is to correct both the $\delta m_h^2$ and $\delta \lambda$ to keep the Golstone massless. However, another equivalent opinion or method is to shift the vev to the ``corrected'' minimum to keep Goldstone boson's massless. In the finite temperature case we will adopt the later standpoint to eliminate the $\Pi_U(k)$ in (\ref{SelfEnergyFormalism}) in our following text. Then we are now ready to discuss the finite temperature case.

Define $u^{\mu} = (1,0,0,0)$ to specify the rest frame of the system, and let $u_{\mu}^T = u_{\mu} - k_{\mu} \frac{u \cdot k}{k^2}$.  $\overline{\Pi}^{\mu \nu}(k)$ can generally be decomposed to the following terms:\cite{ProjectorSource}
\begin{eqnarray}
\overline{\Pi}^{\mu \nu}(k) = \Pi_T(k) P_T^{\mu \nu} + \Pi_L(k) P_L^{\prime \mu \nu} + \Pi_S(k) S^{\mu \nu} + \Pi_U(k) P_U^{\mu \nu},
\end{eqnarray}
where
\begin{eqnarray}
P_L^{\prime \mu \nu} &=& -g^{\mu \nu} + \frac{k^{\mu} k^{\nu}}{k^2} - P_T^{\mu \nu}, \nonumber \\
P_U^{\mu \nu} &=& \frac{k^{\mu} k^{\nu}}{k^2}, \nonumber \\
S^{\mu \nu} &=& \frac{1}{2 |\vec{k}|} (k^{\mu} u^{\nu T} + k^{\nu} u^{\mu T}). \label{GeneralProjectors}
\end{eqnarray}
Notice that we have slightly modified some factors  compared with Ref.~\cite{ProjectorSource} for later practical usage. Then $\overline{\Pi}^{MN}(k)$ can be written as
\begin{eqnarray}
\overline{\Pi}^{MN}(k) = \left( \begin{array}{cc}
\Pi_T(k) P_T^{\mu \nu} + \Pi_L(k) P_L^{\prime \mu \nu} + \Pi_S(k) S^{\mu \nu} + \Pi_U(k) P_U^{\mu \nu} & C k^{\mu} + D u^{T \mu} \\
C^* k^{\nu} + D^* u^{T \nu} & E
\end{array} \right).
\end{eqnarray}
This is the general form to decompose the vector boson's self-energy. We then apply the extended Ward-Takahashi identity in the broken phase $k^*_M \overline{\Pi}^{MN}(k) = 0$ (See Ref.~\cite{ChenjunmouPhysicalGauge, WTIdentityBrokenPhase} for some discussions. We will also discuss this briefly in our Appendix \ref{WTIdentityAppendix}.) to constrain the parameters, where $\frac{k_M^*}{m_A}=(\frac{k_{\mu}}{m_A}, -i)$. Comparing the tensor structures, one can acquire
\begin{eqnarray}
\frac{\Pi_U(k)}{m_A} - C^* i &=& 0, \label{WTEquivalent}\\
\frac{\Pi_S(k) k^2}{2 |\vec{k}| m_A} - D^* i &=& 0, \\
C \frac{k^2}{m_A} - i E &=& 0.	\label{CriticalEquation}
\end{eqnarray}
Here, $C$, $D$, $E$ should all take a $(k)$ dependence which is omitted to be written for brevity. In fact, $E$ corrects the Goldstone boson's mass term. A non zero Goldstone boson's mass term means the departure from the minimum in the new ``heated potential''. Heading for a minimum looks like introducing a counter term to cancel $E$ when $k=0$. Therefore (\ref{WTEquivalent}, \ref{CriticalEquation}) tell us that $\Pi_U(k=0)=0$ in the new minimum. We rewrite the (\ref{GeneralProjectors}) after eliminating $C$, $D$ and $E$,
\begin{eqnarray}
\overline{\Pi}^{MN}(k) = \left( \begin{array}{cc}
\Pi_T(k) P_T^{\mu \nu} + \Pi_L(k) P_L^{\prime \mu \nu} + \Pi_S(k) S^{\mu \nu} + \Pi_U(k) P_U^{\mu \nu} &  -\frac{\Pi_S(k) k^2}{2 i |\vec{k}| m_A} u^{T \mu} + i \Pi_U(k) \frac{k^{\mu}}{m_A} \\
\frac{\Pi_S(k) k^2}{2 i |\vec{k}| m_A} u^{T \nu} - i \Pi_U(k) \frac{k^{\mu}}{m_A}  & \frac{k^2}{m_A^2} \Pi_U(k)
\end{array} \right). \label{SelfEnergyFormalism}
\end{eqnarray}

Notice that $P_L P_T=P_T P_L=P_T S = S P_T = 0 = P^{\prime}_L P_T=P_T P^{\prime}_L=0$, $P_L P^{\prime}_L P_L = P_L$, $P_L S P_L = P_L$,  and when performing the calculations, e.g., the $P_L^{MN} \overline{\Pi}_{MN}(k)$, the ``metric'' $g_{MN} = \text{diag}(1, -1, -1, -1, -1)$ is required. After some tedious calculations of summing over all the ``self-energy strings'', the full propagator (\ref{FullPropagatorProjector}) finally changes to
\begin{eqnarray}
D_{a b}^{\text{full}, MN} (k) = U_{a c}(k) \left( \begin{array}{cc}
D_0^{\text{full}, MN}(k) & 0 \\
0 & D_0^{\text{full} *, MN}(k)
\end{array} \right)_{cd} U_{db}(k),
\end{eqnarray}
where
\begin{eqnarray}
D_0^{\text{full}, MN}(k) &=& \frac{i}{k^2-m_{A}^2 - \Pi_T(k) + i \epsilon} P_T + \frac{i}{k^2-m_{A}^2 - \Pi_L (k) + i \epsilon} P_L \nonumber \\
&+& \frac{1}{1-\frac{\Pi_U(k)}{m_A^2}} \frac{i}{k^2 + i \epsilon} \left[ \begin{array}{cc}
0_{4 \times 4} & 0_{4 \times 1} \\
0_{1 \times 4} & 1
\end{array} \right]. \label{CompletePropagator}
\end{eqnarray}
As in the Ref.~\cite{ProjectorSource}, $\Pi_S$ does not contribute to the mass shifts.

Usually, when one applies the hard thermal loop approximation, $\Pi_L$ and $\Pi_T$ have the universal formats and are given by (Ref.~\cite{Dispersion}, cited on Page 124 of Ref.~\cite{ThermalFieldBook})
\begin{eqnarray}
\Pi_L (k) &=& - \frac{2 m_E^2 k^2}{\vec{k}^2} \left( 1-\frac{k^0}{|\vec{k}|} Q_0 (\frac{k^0}{|\vec{k}|}) \right), \nonumber \\
\Pi_T (k) &=& \frac{1}{2} (2 m_E^2 - \Pi_L(k) ). \label{HTLResult}
\end{eqnarray}
where
\begin{eqnarray}
Q_0 ( \frac{k^0}{|\vec{k}|} ) = \frac{1}{2} \ln \frac{k_0 + |\vec{k}|}{k_0 - |\vec{k}|} ,
\end{eqnarray}
and $m_E$ is the thermal mass parameter depending on the temperature $T$ of the longitudinal polarization calculated in the Euclidean space. The logarithm in this function takes the branch cut connecting the $k_0 = \pm|\vec{k}|$, therefore Im$[Q_0(x+i \epsilon)] = -\frac{i \pi}{2}$ for $|x|<1$ and an infinite small positive $\epsilon$. These arise from the vector boson's coupling with all the particles (including itself in the non-abelian situation). The detailed calculations of the thermal masses are beyond the discussions of this paper. We therefore treat $m_E$ as a parameter.

(\ref{HTLResult}) is equivalent to the effective operator (See Ref.~\cite{VectorThermalL1, VectorThermalL2} for early discussions. Page 185 of Ref.~\cite{MikkoBook} provides the following formalism.)
\begin{eqnarray}
\mathcal{L} \subset \frac{m_E^2}{2} \int d \Omega_v \text{Tr} \left[ \left( \frac{1}{\mathcal{V} \cdot \mathcal{D}} \mathcal{V}^{\alpha} F_{\alpha \mu} \right) \left( \frac{1}{\mathcal{V} \cdot \mathcal{D}} \mathcal{V}^{\beta} F_{\beta}^{\mu}\right)\right], \label{MassCorrect}
\end{eqnarray}
where $\mathcal{V} = (1, \frac{\vec{k}}{k_0})$ is a light-like four-velocity, and $\mathcal{D}$ is the covariant derivative in the adjoint representation.

Usually, the thermal corrections introduce extra imaginary parts in the denominators of the propagators. In this paper, we are trying to figure out the ``on-shell'' behaviours of the vector bosons, so we ignore these extra widths. The polarization vectors of a transverse vector boson remain unchanged, only the dispersion relation changes to $k^2 = m_A^2 + \Pi_T(k)$.  For an on-shell longitudinal vector boson, solve the equation $k^2-m_{A}^2-\Pi_L(k) =0$ to acquire the effective total thermal mass $k^2 = m_{A}^2+\Pi_L(k) = m_{A}^{\prime 2}(k)$ dependent on $k$. From the decomposition relations (\ref{ProjectorL}, \ref{LongitudinalDefinition}, \ref{PLDefinition}) we can acquire the polarization vector to be
\begin{eqnarray}
\epsilon_L^{\prime} = \left( 
\begin{array}{c}
-\frac{m_{A}^{\prime}}{n \cdot k} n^{\mu} \\
i \frac{m_{A}}{m_{A}^{\prime}}
\end{array}
\right). \label{LongitudinalPolarization}
\end{eqnarray}
Generally $m_{A}^{\prime}$ rises up as the temperature arises, so the Goldstone component becomes suppressed by $\frac{m_A}{m_A^{\prime}} < 1$. This implies that the longitudinal polarization of the vector boson is partly ``spewing out'' the once ``eaten'' Goldstone component enforced by the temperature, and is looking more and more likely to become a quasi ``plasmon'' as in the massless vector boson case as we heat the system.

The residue of the vector bosons are also shifted. Define
\begin{eqnarray}
Z_{T, L}(k) = \frac{2 k_0}{2 k_0 - \frac{\partial \Pi_{T, L}(k)}{\partial k_0}}, \label{RenormFactor}
\end{eqnarray}
as the ``wave-function renormalization parameter'', then each external leg of the transverse or longitudinal vector boson should be multiplied with $\sqrt{Z_{T,L}(k)}$.

Now we collect all the terms within the Goldstone component. Since $\Pi_U(k=0) = 0$, and usually $\Pi_U$ changes slowly as $k$ changes, we can ignore the $\Pi_U(k)$ contributions to approximate the $D_0^{\text{full}, 44}$. Summing over all the corresponding elements in the three terms on the right-handed side of the equation (\ref{CompletePropagator}), we finally acquire
\begin{eqnarray}
\Delta^F_{\text{GS}}(k)=\frac{k^2- \Pi_L(k) +i \epsilon}{k^2-m_{A}^{2} - \Pi_L(k) +i \epsilon} \frac{i}{k^2+i \epsilon}. \label{GoldstoneResidue}
\end{eqnarray}
Besides the previously discussed $k^2=m_A^2+\Pi_L(k)$ pole corresponding to the longitudinal polarization mode, it seems that a massless pole $k^2=0$ arises, indicating the appearance of a  scalar degree of freedom. Generally a pole does arise for the complete $\Pi_L(k)$ and non-zero $\Pi_U(k)$ formalisms in the finite temperature environment. However, if we only consider the dominate hard thermal loop contributions in (\ref{HTLResult}), the appearance of $k^2$ in the numerator of the $\Pi_L(k)$ unfortunately cancels this massless pole. In a word, the ``independent'' Goldstone boson disappears at the finite temperature. 

Should the Goldstone degree of freedom completely disappear in the thermal plasma? (\ref{LongitudinalPolarization}) prompts us that only one part of the Goldstone boson had been ``eaten'', then where is the rest part? Eliminating this abruptly will cause an unacceptable discontinuity of degrees of freedom as well as the physical observables before and after the second order phase transition (or more precisely, the ``cross-over''\cite{CrossOver}), because the longitudinal vector boson had no time to take over all the Goldstone boson's legacy when $m_A \sim 0$ is still small just after the cross-over. We will aim at finding back the Goldstone boson's fraction in the next section.

\section{``Quasi-Pole'' Approximation} \label{QuasiPoleSection}

If we carefully examine the analytic performances of (\ref{GoldstoneResidue}) the moment before and after the crossover, when $m_A^2$ start to increase from 0, the $k^2=0$ pole disappears and is replaced by a branch cut $-1 \leq \frac{k^0}{|\vec{k}|} \leq 1$ which fills all of the (phase velocity) tachyonic area.

\begin{figure}
\includegraphics[width=0.48\textwidth]{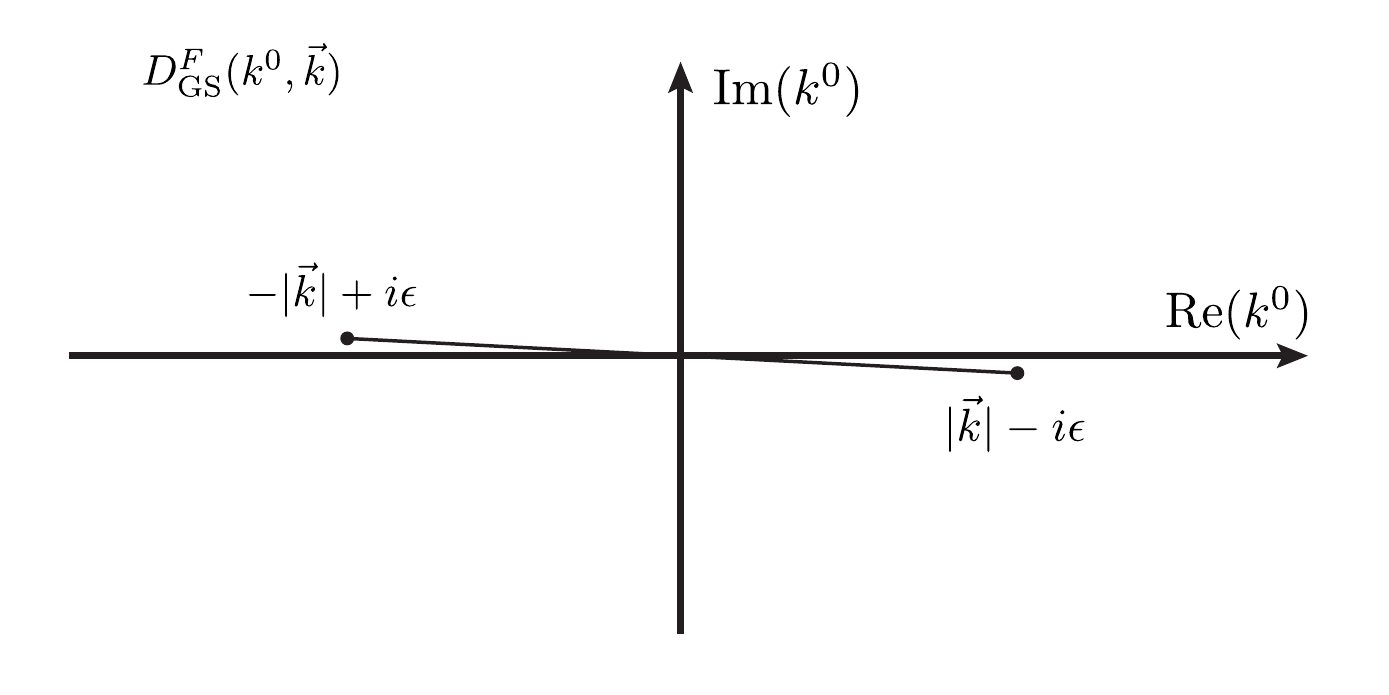}
\includegraphics[width=0.48\textwidth]{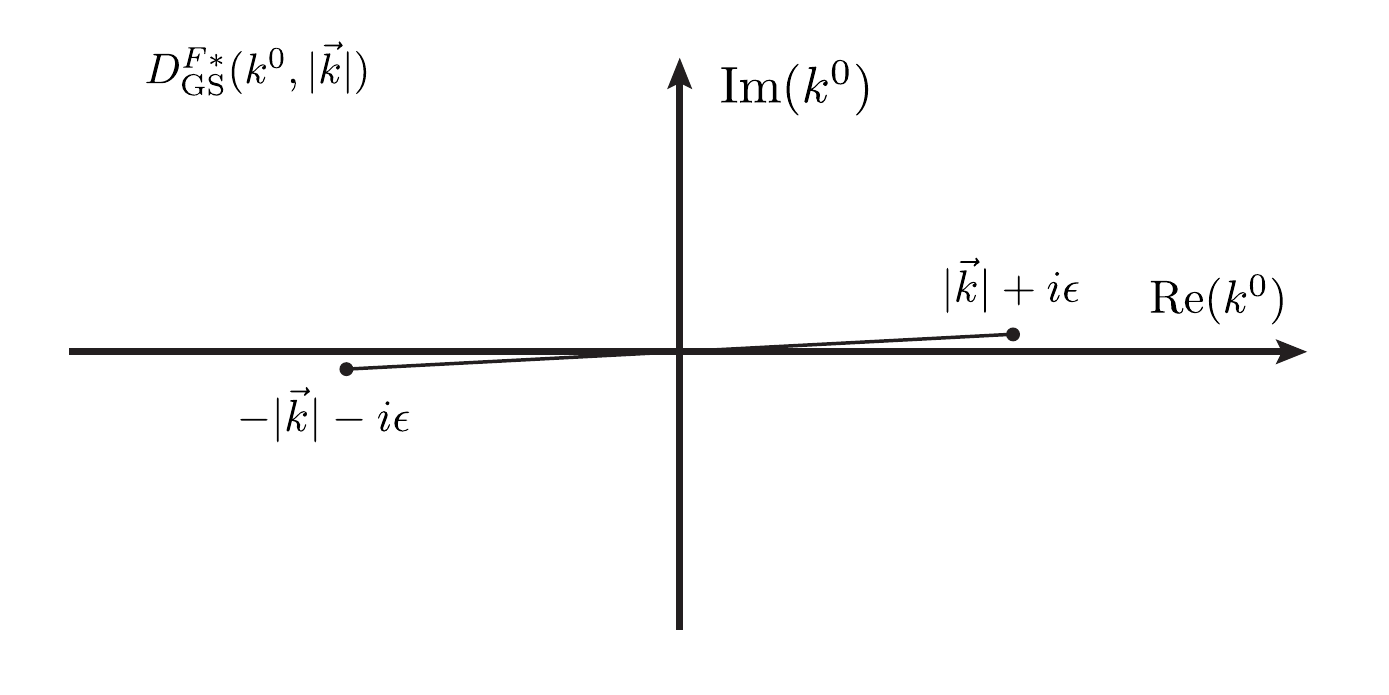}
\caption{Branch points and branch cuts of the $D_{\text{GS}}^{F(*)}(k^0, |\vec{k}|)$. Notice that $D_{\text{GS}}^{F*}(k^0, |\vec{k}|)$ is not the complex conjugation of $D_{\text{GS}}^{F}(k^0, |\vec{k}|)$ through all over the complex plane. Only the branch points and branch cut are different, so equation $D_{\text{GS}}^{F(*)}(k^0, |\vec{k}|)=(D_{\text{GS}}^{F}(k^0, |\vec{k}|))^*$ is restricted on the real axis.\label{BranchStructure}}
\end{figure}

The usual inclusive sum rules method sometimes adopts the Keldysh, or r/a basis to avoid the intersect between the $k^0$ integration contours and the branching cuts (See an introduction on page 173 in Ref.~\cite{MikkoBook}). Abandoning this aesthetic selection, we choose the more intuitive however equivalent formalism of the general ``$\sigma$ choice''. The tree-level propagators are shown on Page 53 of Ref.~\cite{ThermalEarlyReview1}, and only isolated poles were manipulated there. By writing down the ``spectrum representation'' of a branch cut,  we can replace the tree-level $\frac{i}{k^2-m^2+ i \epsilon}$ and $2 \pi \delta(k^2-m^2)$ with an integration of the ``aligned poles''. That is to say,
\begin{eqnarray}
\Delta_{\text{GS}}^{F}(k) \sim \int \frac{d k^{0 \prime}}{2 \pi} \rho_F(k^{0 \prime}, |\vec{k}|) \frac{i}{k^0-k^{0 \prime}} + \text{iso.~pole~cont.},
\end{eqnarray}
where $\rho_F(k^{0 \prime}, |\vec{k}|) = -2 \text{Im}[i \Delta^F_{\text{GS}}(k^0+i \epsilon, \vec{k})]$, and we did not explicitly write down the familiar non-tachyonic isolated pole contributions (iso.~pole~cont.). Considering the imaginary part in the real axis, all of the $\frac{i}{k^2-m^2 \pm i \epsilon}$ in the tree-level propagator should be replaced with $\frac{i \rho_F(k^{0 \prime}, |\vec{k}|)}{k^0-k^{0 \prime} + i k^{0 \prime} \epsilon} $, and noticing that $\rho_F(k^{0 \prime}, |\vec{k}|)$ is an odd function on $k^{0 \prime}$, and it is easy to prove that $\rho_F(0<k^{0 \prime}<|\vec{k}|, |\vec{k}|) <=0$, then $2 \pi \delta(k^2-m^2)$ should also be replaced with $ |\rho_F(k^{0 \prime}, |\vec{k}|)| 2 \pi \delta(k^0 - k^{0 \prime}) = i \rho_F(k^{0 \prime}, |\vec{k}|) \left( \frac{1}{k^0-k^{0 \prime} - i k^{0 \prime} \epsilon} - \frac{1}{k^0 - k^{0 \prime} + i k^{0\prime} \epsilon} \right)$. After performing the integration $\int \frac{d k^0}{2 \pi}$, we finally learn that the definition of $\Delta_{\text{GS}}^F$ should become
\begin{eqnarray}
\Delta_{\text{GS}}^{F}(k) = \int \frac{d k^{0 \prime}}{2 \pi} \rho_F(k^{0 \prime}, |\vec{k}|) \frac{i}{k^0-k^{0 \prime} + i k^{0 \prime} \epsilon} + \text{iso.~pole~cont.},
\end{eqnarray}
so the branch cut should span from $k^0= -|\vec{k}|+i \epsilon$ to $k^0 = |\vec{k}|-i \epsilon$, and just goes through the origin. If we restrict our aspect on the real axis function values, the ``complex conjugation'' $D_{\text{GS}}^{F*}$ just takes the opposite branch cut path from $k^0= -|\vec{k}|-i \epsilon$ to $k^0 = |\vec{k}|+i \epsilon$ (See Fig.~\ref{BranchStructure} for the branch points and branch cuts of $D_{\text{GS}}^{F(*)}$), and
\begin{eqnarray}
\Delta_{\text{GS}}^{F*}(k) = \int \frac{d k^{0 \prime}}{2 \pi} \rho_F(k^{0 \prime}, |\vec{k}|) \frac{-i}{k^0-k^{0 \prime} - i k^{0 \prime} \epsilon} + \text{iso.~pole~cont.}.
\end{eqnarray}
Therefore,
\begin{eqnarray}
\frac{\Delta_{\text{GS}}^{F}(k) - \Delta_{\text{GS}}^{F*}(k)}{2} = -|\rho_F (k^0, \vec{k})| + \text{iso.~pole~cont.}.
\end{eqnarray}
We can then write the explicit expressions of the $\sigma$-dependent $\Delta_{\text{GS}, 11, 12, 21, 22}^F$,
\begin{eqnarray}
& \Delta_{\text{GS}, 11}^F(k) &= \Delta_{\text{GS}}(k) - n(k^0) \frac{\Delta_{\text{GS}}^F(k) - \Delta_{\text{GS}}^{F*}(k)}{2} = \Delta_{\text{GS}, 22}^{F*}(k), \nonumber \\
& \Delta_{\text{GS}, 12}^F(k) &= e^{\sigma k^0} [n(k^0) + \theta(-k^0)] \frac{\Delta_{\text{GS}}^F(k) - \Delta_{\text{GS}}^{F*}(k)}{2}, \nonumber \\
& \Delta_{\text{GS}, 21}^F(k) &= e^{-\sigma k^0} [n(k^0) + \theta(k^0)] \frac{\Delta_{\text{GS}}^F(k) - \Delta_{\text{GS}}^{F*}(k)}{2}.
\end{eqnarray}

By principle, when we calculate the imaginary part of the inclusive $\Pi^{>}$ or $\Pi^{<}$ functions for the physical observables in the real-time formalism, our integration along the real axis of $k^0$ encounters the branch cut of the $D_{\text{GS}}^{F(*)}$ and inevitably intersect with it as shown in Fig.~\ref{BranchStructure}. The imaginary part is difficult to be integrated analytically, however, usually we find that the Im$[i \Delta_{\text{GS}}^F(0, |\vec{k}|)]$ peaks in the vicinity of the branch points $k^0 = \pm |\vec{k}|$ if we plot their values, and the they become zero when $k^0 = 0$ since it is an odd function. These peaks becomes especially evident when $m_A^2 \ll m_E^2$, and such ``branch cuts'' extremely resemble the $k^2=0$, or $k^0 = \pm | \vec{k}|$ poles. This is plausible, and it is obvious that these branch cuts are the inheritors of the disappearing Goldstone poles. In this paper, we call such a branch cut a pair of ``quasi-poles'', which are found to be the fragments of the disappeared Goldstone boson.

\begin{figure}
\includegraphics[width=6in]{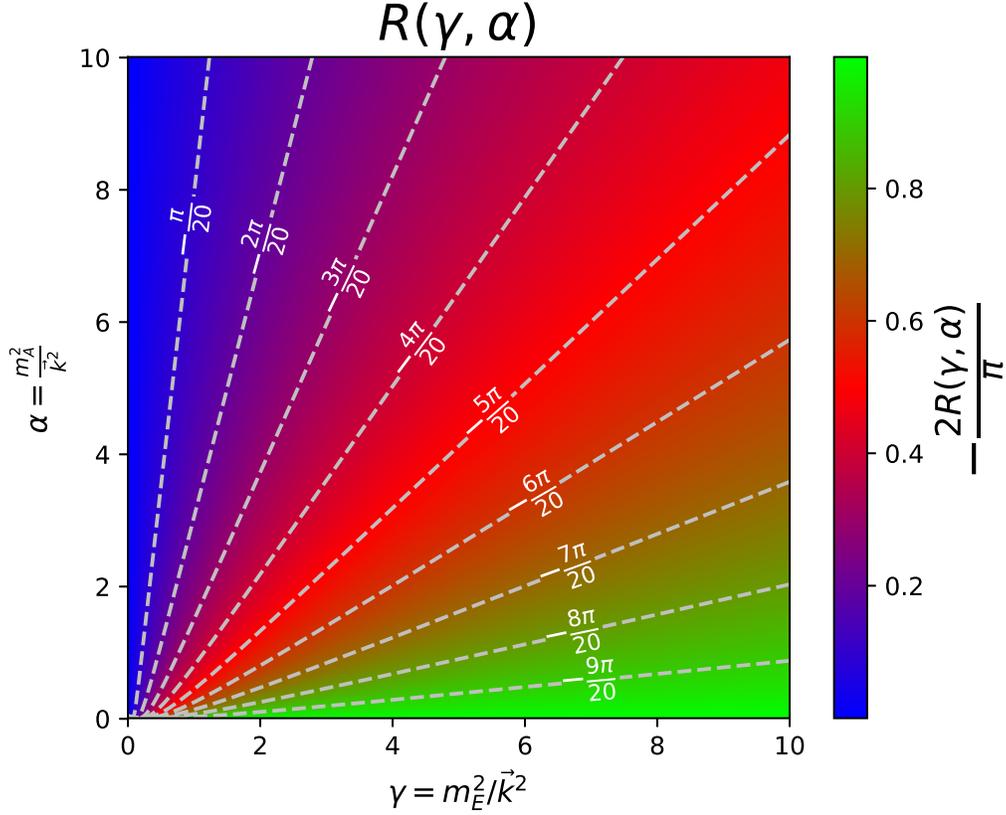}
\caption{$R(\gamma, \alpha)$, or $Z_{\rm{GS}}$ values as a function of $\alpha$ and $\gamma$. \label{RValues}}
\end{figure}

As an approximation, we can replace the two halves of the branch cut $-|\vec{k}|\leq k^0<0$ and $0< k^0 \leq |\vec{k}|$ with two poles $k^0 = \pm (|\vec{k}|-i \epsilon)$. All of the branch cut's imaginary part are collected and appointed to become the pole residues
\begin{eqnarray}
\text{Res}[k^0=\pm(|\vec{k}| - i \epsilon)] = \int_0^{|\vec{k}|+\delta} -2 \text{Im}[ i \Delta^F_{\text{GS}}(k^0, \vec{k}) ] dk^0, 
\end{eqnarray}
where $\delta$ is a small positive value satisfying $\epsilon \ll \delta \ll m_A$ to collect all of the imaginary parts near the branch cut while keep the neighbour of the longitudinal vector boson's pole untouched. Define
\begin{eqnarray}
x = \frac{k^0}{|\vec{k}|},~~\gamma=\frac{m_E^2}{\vec{k}^2},~~\alpha=\frac{m_A^2}{\vec{k}^2},
\end{eqnarray}
then
\begin{eqnarray}
& & \int_0^{|\vec{k}|+\delta}-\text{Im} [i \Delta_{\text{GS}}^F (k^0, \vec{k})] dk^0 \nonumber \\
&=& \frac{1}{\vec{k}}\int_0^{1+\delta} \text{Im} \left[ \frac{x^2-1+ 2 \gamma (x^2-1+i \epsilon)  (1-x Q_0(x))} {x^2-1+i \epsilon-\alpha + 2 \gamma (x^2-1)  (1-x Q_0(x))+i \epsilon} \frac{1}{x^2-1+i \epsilon} \right] dx \nonumber \\
&\overset{\Delta}{=}& \frac{1}{\vec{k}} R(\gamma, \alpha). \label{RDefinition}
\end{eqnarray}
$R(\gamma, \alpha)$ is a two-dimensional function that can be calculated numerically. Its values are plotted in Fig.~\ref{RValues}. Notice that $R(\gamma \neq 0, \alpha \rightarrow 0) = -\frac{\pi}{2}$, indicating the recovery of a massless Goldstone pole during the crossover moment. As the vev arises and increases, $m_A$ increases and $R(\gamma, \alpha)$ decreases, implying the gradual disappearance of the Goldstone boson. At the same time, the Goldstone component of the longitudinal mode of the vector boson increases, just as we have discussed in the previous section.

We have also tried to use the following formula to fit the $R(\gamma, \alpha)$:
\begin{eqnarray}
R(\gamma, \alpha) \simeq -\frac{\pi}{2}+A \left( \frac{e^P - e^{-P}}{e^P + e^{-P}} \right)^F,
\end{eqnarray}
where
\begin{eqnarray}
& & P=\frac{\alpha^{B+\frac{G}{\gamma^H + I}}}{C \gamma^D+E}, \nonumber \\
& & A=1.5339~B=0.16484,~C=0.47210,~D=0.20252~E=2.5680 \times 10^{-22}, \nonumber \\
& & F=15.64287469,~G=4.82049 \times 10^{-4},~H=0.26394~,I=5.15737 \times 10^{-3},
\end{eqnarray}
are the best fitted parameters that we have acquired.

With this quasi-pole approximation, we can regard the ``tachyonic'' branch cut as a massless Goldstone boson, with the ``wave function renormalization parameter'' 
\begin{eqnarray}
Z_{\text{GS}} = -\frac{2 R(\gamma, \alpha)}{\pi},
\end{eqnarray}
which always $\leq$1.

Finally, we are ready to write down the Feynmann rules of the originally massive vector bosons as the external legs by the following steps
\begin{itemize}
\item Calculate the effective thermal potential of the Higgs boson as usual, find out the vev $v$ for the minimum, then calculate the ``original mass'' of the vector boson $m_A$.
\item Calculate the $\Pi_L(k)$ and $\Pi_T(k)$. For the hard thermal loop approximation, these can be attributed to calculating the thermal mass for the zero-energy longitudinal vector boson $m_E$ as usual. Remember to calculate $Z_{T,L}(k)$ from (\ref{RenormFactor}) as well.
\item For the vector boson/Goldstone inner propagators, directly use (\ref{CompletePropagator}).
\item For an external leg of a vector boson, the transverse polarization is the same as the zero-temperature situation. Notice that the on-shell dispersion relation should be modified to $k^2 = m_A^2 + \Pi_T(k)$. This usually involves solving the transcendental equations. A factor of $\sqrt{Z_T(k)}$ is also required.
\item For an external leg of a longitudinal vector boson, the on-shell relation is $k^2 = m_A^2 + \Pi_L(k)$. Calculate $m_A^{\prime 2} = m_A^2 + \Pi_L(k)$, then the polarization vector should be the form of (\ref{LongitudinalPolarization}). A factor of $\sqrt{Z_L(k)}$ is required as well.
\item External legs of the Goldstone boson should not be forgotten. There should be a $\sqrt{Z_{\text{GS}}}$ factor for each of the external Goldstone bosons.
\item Notice for each external leg, a factor $\sqrt{n_B(k)}$ is sometimes required for each initial state vector boson (or Goldstone boson), and a factor $\sqrt{1+n_B(k)}$ is always required for each final state vector boson (or Goldstone boson).
\end{itemize}

\section{Equivalence in Other Gauges}

The physical gauges are not the main stream of the practical calculations in the literature. In the following text, we discuss how to acquire the similar result in the $R_\xi$ gauge. Note that the cancellation of the $\xi$-depencence in computing the physical observables is currently beyond our ability. However, in the HTL approximation, all the $\xi$ dependence is attributed to the effective potential. If all the diagrams are resummed, the $\xi$-dependence of observables are expected to be killed, and the basic formalisms of our following discussions are also expected to retain.

In the $R_\xi$ gauge, the $5 \times 5$ propagators like the (\ref{FullPropagator})  are partly diagonalized to eliminate the $\phi$-$A^{\mu}$ ``crossing-terms''.  The Goldstone part then becomes $\frac{k^2-m_A^2}{k^2- \xi m_A^2}$. Commonly, the vev of the Higgs boson is explicitly encoded in the gauge fixed effective Lagrangian, which appears to be inconvenient when the vev varies as the temperature changes. In Ref.~\cite{GaugeFixingTerm}, the authors suggect such a gauge fixing term
\begin{eqnarray}
& & \mathcal{F} = \partial_{\mu} A^{\mu} + \frac{i}{2} \xi g (H^2 - H^{\dagger 2}), \nonumber \\
& & \mathcal{L}_{\text{G.F.}} = \frac{1}{2 \xi} \mathcal{F}^2,
\end{eqnarray}
to automatically adjust the vevs in different temperatures. Here (\ref{FullPropagator}) is replaced with
\begin{eqnarray}
\langle (A^{\mu}, \phi), (A_{\mu}, \phi) \rangle = \frac{i}{k^2-m_A^2+i \epsilon} \left(
\begin{array}{cc}
-(g^{\mu \nu} - \frac{k^{\mu} k^{\nu} (1-\xi) }{k^2 - \xi m_A^2+ i \epsilon} ) & 0_{4 \times 1} \\
0_{1 \times 4} & \frac{k^2 - m_A^2 +i \epsilon}{k^2 - \xi m_A^2 + i \epsilon}
\end{array} 
 \right).
\end{eqnarray}
This tree-level propagator is decomposed to
\begin{eqnarray}
\langle (A^{\mu}, \phi), (A^{\nu}, \phi) \rangle = \frac{i}{k^2-m_{A}^2+i \epsilon} (P_T + P_L^{\prime} + P_{R\text{-}\xi}),
\end{eqnarray}
where the $5 \times 5$ longitudinal part $P_L^{\prime}$ for this situation is extended directly from the $4 \times 4$ $P_L^{\prime}$ defined in (\ref{GeneralProjectors}), and
\begin{eqnarray}
P_{R\text{-}\xi} = \left( \begin{array}{cc}
\left[ \frac{(1-\xi)}{k^2 - \xi m_A^2 + i \epsilon} - \frac{1}{k^2+ i \epsilon} \right] k^{\mu} k^{\nu} & 0_{4 \times 1} \\
0_{1 \times 4} & \frac{k^2-m_A^2+ i \epsilon}{k^2 - \xi m_A^2 + i \epsilon}
\end{array}\right) = \left( \begin{array}{cc}
\frac{-\xi(k^2 - m_A^2 + i \epsilon)}{(k^2 - \xi m_A^2 + i \epsilon) (k^2 + i \epsilon)} k^{\mu} k^{\nu} & 0_{4 \times 1} \\
0_{1 \times 4} & \frac{k^2-m_A^2+ i \epsilon}{k^2 - \xi m_A^2 + i \epsilon}
\end{array}\right).
\end{eqnarray}
In this case, the pole structure is much more complicated than the gauge equivalence gauge. In the zero temperature, the Ward-Takahashi identity guarantees that $\xi$-dependent pole $k^2 = \xi m_A^2$ cancels within the gauge and Goldstone parts. the $k^2 = 0$ pole immediately disappears by summing the $P_L^{\prime} + P_{R\text{-}\xi}$. Then, again resumming all the ``self-energy strings'' in one-loop level involving (\ref{SelfEnergyFormalism}) as in the case of the Goldstone-equivalence gauge, we acquire
\begin{eqnarray}
D_0^{\text{full}, MN}(k) &=& \frac{i}{k^2-m_{A}^2 - \Pi_T(k) + i \epsilon} P_T + \frac{i}{k^2-m_{A}^2 - \Pi_L (k) + i \epsilon} P_L^{\prime} \nonumber \\
&+& \frac{i}{k^2- m_A^2 + i \epsilon} P_{R\text{-}\xi} + \frac{\Pi_U(k)}{1-\frac{\Pi_U(k)}{m_A^2}} \frac{i}{(k^2-\xi m_A^2 + i \epsilon)^2} P_{R\text{-}\xi}^{\text{Loop}}, \label{RXiCompletePropagator}
\end{eqnarray}
where
\begin{eqnarray}
P_{R\text{-}\xi}^{\text{Loop}} = \left( \begin{array}{cc}
\frac{\xi^2 k^{\mu} k^{\nu}}{k^2} & i \frac{\xi k^{\nu}}{m_A} \\
-i \frac{\xi k^{\mu}}{m_A} & \frac{k^2}{m_A^2}
\end{array} \right).
\end{eqnarray}
One can compare our complete $R_\xi$ results with the corresponding formulas in Appendix B, Ref.~\cite{BelowTcLaine2}. Additional $\xi$-dependent terms arise in our paper and the structure of the propagators will be studied in more detail.

Besides the $k^2 = \xi m^2$ poles in the $P_{R\text{-}\xi}$ elements which are destined to be cancelled due to the extended Ward-Takahashi identity in the physical observable calculations, $P_{R\text{-}\xi}^{\text{Loop}}$ introduces the annoying $\frac{1}{(k^2 - \xi m^2)^2}$ double poles. Fortunately, notice that 
\begin{eqnarray}
P_{R\text{-}\xi, MN}^{\text{Loop}} &=& (\frac{\xi}{\sqrt{k^2}} k^{\mu}, i \frac{\sqrt{k^2}}{m_A})_M (\frac{\xi}{\sqrt{k^2}} k^{\nu}, -i\frac{\sqrt{k^2}}{m_A})_N  \nonumber \\
&=& \frac{\xi}{\sqrt{k^2}} k_M k_N^* + \frac{\xi}{\sqrt{k^2}} (k_M t_N^* + t_M k_N^*) + t_M t_N^*, \label{PLoopDecompose}
\end{eqnarray}
where $t_N = (0, 0, 0, i \frac{k^2 - \xi m_A^2}{m_A \sqrt{k^2}})$. After contracting the indices $M$, $N$ with other parts of the diagram or with the external polarization vectors, the Ward-Takahashi identity will straightforwardly kill the first and second term of the (\ref{PLoopDecompose}), and the double poles are then cancelled by the $(k^2 - \xi m_A^2)^2$ term arising from the third term. Therefore, these double poles are proved to be non-physical.

Then let us collect all the contributions to the terms proportional to $\frac{i}{k^2 + i \epsilon}$,
\begin{eqnarray}
\frac{i k^{\mu} k^{\nu}}{k^2 + i \epsilon} \left[ \frac{1}{m_A^2} - \frac{1}{m_A^2 + \Pi_L(k)} + \frac{\Pi_U(k)}{1-\frac{\Pi_U(k)}{m_A^2}} \frac{1}{m_A^4}  \right]. \label{RXiGoldstone} \label{MasslessGoldstoneRxi1}
\end{eqnarray}
This induces the ``polarization vector'' $\epsilon^{\mu}$ $\propto k^{\mu}$ contributions. However, the Ward-Takahashi identity can replace such an external vector boson by a scalar mode with a Yukawa-like coupling, and these two modes are undistinguishable. contracting the $k^{\mu} k^{\nu}$ in (\ref{MasslessGoldstoneRxi1}) with other (parts of the) diagrams, and apply the Ward-Identity replace the $k^{\mu} k^{\nu}$ by $m_A^2$. Finally, (\ref{MasslessGoldstoneRxi1}) becomes
\begin{eqnarray}
& & \frac{i m_A^2}{k^2 + i \epsilon} \left[ \frac{1}{m_A^2} - \frac{1}{m_A^2 + \Pi_L(k)} + \frac{\Pi_U(k)}{1-\frac{\Pi_U(k)}{m_A^2}} \frac{1}{m_A^4}  \right]. \nonumber \\
&=& \frac{i}{k^2 + i \epsilon}\left[ \frac{1}{1-\frac{m_A^2}{\Pi_U(k)}} + \frac{\Pi_L(k)}{m_A^2 + \Pi_L(k)} \right], \label{MasslessGoldstoneRxi2}
\end{eqnarray} 
which is exactly the same format with the full factor of the $\frac{1}{k^2+i \epsilon}$ terms in (\ref{RXiCompletePropagator}). Therefore we reproduce the (\ref{GoldstoneResidue}).  We then omit the rest of the calculations and the similar results that one can acquire.

In the $R_\xi$ gauge, longitudinal polarization vector recovers to the usual $\epsilon_{LU \mu} (k) = (|\vec{k}|, k_0 \frac{\vec{k}}{|\vec{k}|})/\sqrt{k^2}$, with no additional Goldstone part involved. Notice that
\begin{eqnarray}
\epsilon_{L}^M + i \frac{k^M}{m_A^{\prime}} = \left( \begin{array}{c} 
\epsilon_{LU}^{\mu} \\
0
\end{array} \right)^M,	\label{LandauLongitudinalPolarizationDerivation}
\end{eqnarray}
where $\frac{k^M}{m_A}=(\frac{k^{\mu}}{m_A}, -i)$. The extended Ward-Identity guarantees that $k^M$ does not contribute to the amplitude, therefore $\epsilon_L$ and $\epsilon_{LU}$ are equivalent in the practical calculations. However, $\epsilon_{LU}$ conceals the ``eaten'' Goldstone boson contributions, disconcerting us whether it is a real elementary particle, or a quasi-particle ``plasmon'' before we complete all the above analyses.

Although we have finished the $R_\xi$ discussions within several pages, we have to point out that the detailed derivations are much more complicated than the Goldstone equivalence gauge. We omitted some of the cumbersome formula calculations, and warped the common processes for all gauges. Our calculation processes in the $R_\xi$ gauge were also enlightened by the priory knowledges we acquired for the Goldstone equivalence gauge. Without these, even if one might be able to ``guess out'' the correct formalism at the $R_\xi$ gauge before they are validated, however it is still an obstacle for one to clarify the Goldstone boson and longitudinal polarization relationships. This is one of the reasons that we begin this paper with the Goldstone equivalence gauge. From (\ref{RXiGoldstone}), we can see in the $R_\xi$ gauge, the Goldstone branch cut is scattered in all of the $4 \times 4$ propagator elements, making it hard to perform an awkward inclusive calculation by sum rule methods without a felicitous decomposition. In contrast, the physical gauge attributes all the Goldstone contributions to one single element. Equipped with (\ref{PLoopDecompose}, \ref{RXiGoldstone}), the traditional $R_\xi$ gauge inclusive calculations also become well-facilitated.

Another advantage of the Goldstone equivalence gauge compared with the $R_\xi$ gauge is that a continuous (although not necessarily differentiable) ``tree-level'' production/decay rates around the cross-over temperature can be acquired automatically, since in this gauge the Goldstone boson's contributions are attributed to one single component of the polarization vector. On the contrary, in the $R_\xi$ case, when we are jumping over the critical temperature, a continuous matching between the $k^{\mu} k^{\nu}$ induced terms and the Goldstone boson contributions requires an additional one-loop vertex correction.

Coulomb gauge is also a common selection in the literature. In this case, the gauge vector is chosen to be
\begin{eqnarray}
n_C^{\mu} = (0, \vec{k}).
\end{eqnarray}
The longitudinal polarization vector in (\ref{LongitudinalDefinition}) then changes to
\begin{eqnarray}
\epsilon_{LC}^M(k) = \frac{1}{\sqrt{\frac{m_A^2}{k^2} - \frac{n_C^2 m_A^2}{(n_C \cdot k)^2}}} \left(\begin{array}{c}
-\frac{m_A}{n_C \cdot k} (n_C^{\mu} - \frac{n_C^2 k^{\mu}}{n_C \cdot k}) \\
i (\frac{m_A^2}{k^2} - \frac{n_C^2 m_A^2}{(n_C \cdot k)^2})
\end{array} \right),
\end{eqnarray}
and $P_L$ in (\ref{ProjectorL}) should be replaced with
\begin{eqnarray}
P_{LC}^{MN} &=&  \epsilon_{LC}^M \epsilon_{LC}^{N*}.
\end{eqnarray}
The following processes are similar, however it is much more complicated to acquire the (\ref{HTLResult}) then the Goldstone equivalence gauge. Finally, the result is the same as in the Goldstone equivalence gauge.

\section{Gauge Boson Mixing}

In the standard model, the above discussions can be 	directly applied on $W^{\pm}$ bosons. The $Z$ and $\gamma$ bosons arise through the mixing between the $W^3_{\mu}$ and the hyper-charge gauge boson $B^{\mu}$. Thermal mass might disturb the mixing angle, and add inconvenience on the Goldstone boson sector. Therefore, based upon a $U(1) \times U(1)$ toy model, we briefly illustrate the thermal effects of the mixed gauge vector boson. Many details in deriving the following formulas are similar to Sec.~\ref{ThermalAdded}, so we eliminate these details in this section.

Based upon (\ref{LOrigin}), we introduce another $U(1)_B$ gauge boson $B^{\mu}$ (Do not be confused with the hypercharge gauge field $B^{\mu}$!). Its field strength tenser is defined to be $B_{\mu \nu} = \partial_{\mu} B_{\nu} - \partial_{\nu} B_{\mu}$. The $U(1)$ group corresponding to the original gauge boson $A^{\mu}$ is renamed with $U(1)_A$ to avoid the confusions. The toy-model Lagrangian becomes
\begin{eqnarray}
\mathcal{L} \supset -\frac{1}{4} F_{\mu \nu} F^{\mu \nu} -\frac{1}{4} B_{\mu \nu} B^{\mu \nu} + \mathcal{L}_{\rm{Higgs}}.
\end{eqnarray}
$\mathcal{L}_{\rm{Higgs}}$ includes the Higgs boson(s) taking either or both of the $U(1)_A$ or $U(1)_B$ charges, giving rise to the symmetric mass matrix $M_{V}^2$ of the gauge bosons after acquiring the vevs,
\begin{eqnarray}
\mathcal{L} \supset - \frac{1}{2}\left( A_{\mu} ~~ B_{\mu} \right) M_{V}^2 \left( \begin{array}{c}
A^{\mu} \\
B^{\mu} \end{array} \right) = - \frac{1}{2} \left( A_{\mu} ~~ B_{\mu} \right) \left( \begin{array}{cc}
m_{A}^2 &m_{AB}^2 \\
m_{AB}^2 & m_{B}^2
\end{array} \right) \left( \begin{array}{c}
A^{\mu} \\
B^{\mu} \end{array} \right).
\end{eqnarray}
Thermal mass terms $\Pi_{L,T}(k)$ are also replaced with the formalism of $2 \times 2$ symmetric matrix
\begin{eqnarray}
\Pi_{P}(k) = \left( \begin{array}{cc}
\Pi_P^A(k) & \Pi_P^{AB}(k) \\
\Pi_P^{AB}(k) & \Pi_P^B(k)
\end{array} \right),
\end{eqnarray}
where $P=L,T$. In the HTL approximation, $\Pi_{L,T}^{A,AB,B}(k)$ are still the same formalism with (\ref{HTLResult}), and $m_E$ in the (\ref{HTLResult}) can be parametrized and replaced with $m_E^A$, $m_E^{AB}$, $m_E^{B}$ respectively.

$M_{V}$ and $\Pi_P$ are not necessarily proportional to each other, so they generally are not able to be diagonalized simultaneously. Since the mass terms $\Pi_{L,T}$ depends on the momentum, the diagonalized propagator structures are also damaged. The familiar terms of the propagators $\frac{i}{k^2-m_i^2}$ like (\ref{CompletePropagator}) should be generalized into the matrix form like
\begin{eqnarray}
D_0(k) = i (k^2 I - M_{V}^2 + i \epsilon I)^{-1},
\end{eqnarray}
where $I$ is the identity matrix, and $D_0(k)$ is the propagator matrix. This  ``Secular equation'' $\rm{Det} (k^2 I - M_{V}^2)  = 0$ defines the poles corresponding to the physical particle states. Without loss of generality, we assume that $M_{V}^2$ had already been diagonalized so $m_{AB}^2=0$ in our following discussions. After summing over the cumbersome self energy trails, again eliminating the unimportant $\Pi_U(k)$ terms, the two-gauge boson version of (\ref{CompletePropagator}) becomes
\begin{eqnarray}
D_0^{\rm{full}, QR, MN} (k) &=& i P_T^{MN} \cdot \left\lbrace \left[ k^2 I -M^2_{AB} - \Pi_T(k) + i \epsilon I \right]^{-1} \right\rbrace^{QR} \nonumber \\
&+& i P_L^{QR, MN} \cdot \left\lbrace \left[ k^2 I -M^2_{AB} - \Pi_L(k) + i \epsilon I \right]^{-1} \right\rbrace^{QR} \nonumber \\
&+& \delta^{QR} \delta^{M4} \delta^{N4} \frac{i}{k^2 + i \epsilon}, \label{MixingPropagator}
\end{eqnarray}
where $M$, $N=$0, 1, 2, 3, 4 (4 corresponds to the Goldstone boson), and $Q$, $R$=$A$, $B$. In the second term in (\ref{MixingPropagator}), the Einstein's summation rule does not work on indices $Q$, $R$. $P_T$ had already been defined in (\ref{ProjectorT}), and $P_L^{QR}$ is defined with
\begin{eqnarray}
P_L^{QR, MN} = \epsilon_L^{Q,M} \epsilon_L^{R,N},
\end{eqnarray}
where
\begin{eqnarray}
\epsilon_L^{Q,M} = \left( \begin{array}{c}
-\frac{\sqrt{k^2}}{n \cdot k} n^{\mu} \\
i \frac{m_Q}{\sqrt{k^2}} \end{array} \right).
\end{eqnarray}

For the ``on-shell'' transverse and longitudinal particle momentums, one can solve equations $\rm{Det}[k^2 I -M^2_{AB} - \Pi_T(k)]=0$ and $\rm{Det}[k^2 I -M^2_{AB} - \Pi_L(k)]=0$ for the allowed $k$, and take $k$ into the $[M^2_{AB} - \Pi_T(k)] x=k^2 x$ and $[M^2_{AB} - \Pi_L(k)]=k^2 x$ to solve the vector boson's eigenvector $x$ for the mixing angles at momentum $k$. Notice that the mixing patterns generally depends on momentum $k$, so no universal separated propagators can be written. The on-shell transverse mode's polarization vectors are exactly the same as before, while the longitudinal ones take some subtleties. We suggest extend the $\epsilon_L$ with two Goldstone components,
\begin{eqnarray}
\epsilon_L^M = \left( \begin{array}{c}
-\frac{\sqrt{k^2}}{n \cdot k} n^{\mu} \\
i \frac{x_A m_A}{\sqrt{k^2}} \\
i \frac{x_B m_B}{\sqrt{k^2}}
\end{array} \right),
\end{eqnarray}
where $x=(x_A, x_B)$ is the normalized eigenvector satisfying $[M^2_{AB} - \Pi_L(k)]x=k^2 x$ for this mode in momentum $k$.

For the most important remained Goldstone boson's fraction, we collect all of the Goldstone parts of the $D_0^{\rm{full}}$,
\begin{eqnarray}
D_0^{\rm{full},QR,44} = \frac{i}{k^2+i \epsilon} \sqrt{M_{V}^2}^{QS} \left\lbrace \left[ k^2 I -M^2_{AB} - \Pi_L(k) + i \epsilon \right]^{-1} \right\rbrace^{ST} \sqrt{M_{V}^2}^{TR}+ \frac{i}{k^2} \delta^{QR}, \label{GoldstoneMixingFull}
\end{eqnarray}
where $\sqrt{M_V^2} = \rm{diag}[m_A, m_B]$. One can verify that at high temperature limit when $\Pi_L \rightarrow \infty$, $\frac{i}{k^2}$ recovers in (\ref{GoldstoneMixingFull}) and the massless Goldstone bosons resurrect. In the zero temperature case when $\Pi_L=0$, Goldstone modes are completely ``devoured'' by the vector bosons. Between them, there should be a smooth intermediate state with complicated tensor and analytical structures of $D_0^{\rm{full},QR,44}$, marked by tangles of shaggy poles and branch cuts. One might find it difficult to organize a ``Quasi-pole approximation'' as in Sec.~\ref{QuasiPoleSection} to write down the external Goldstone boson's feynman rules. However, fortunately, when $m_B=0$, which is exactly the situation of the $\gamma$-$Z$ system in the standard model at any temperature, only $D_0^{\rm{full},AA,44}$ remains nonzero and can be calculated analytically.
\begin{eqnarray}
& & D_0^{\rm{full},AA,44}(k) \nonumber \\
&=& \frac{i}{(k^2+i \epsilon)} \frac{m_A^2 (k^2 - \Pi_L^B + i \epsilon )}{(k^2 - m_A^2 - \Pi_L^A + i \epsilon)(k^2 - \Pi_L^B + i \epsilon) - (\Pi_L^{AB} )^2} + \frac{i}{k^2 + i \epsilon}.
\end{eqnarray}
Earnest analysis can still show that the $k^2=0$ poles had been replaced by a branch cut connecting $k^0 = \pm |\vec{k}|$. Replace the $\Delta_{\rm{GS}}^F$ with $D_0^{\rm{full},AA,44}(k)$ in (\ref{RDefinition}), one can repeat the integrations to calculate the $R(\gamma, \alpha)$, finally gathering all the elements to write down the Feynman rules.

\section{Discussions and Possible Applications}

Besides the Goldstone degree of freedom, both the transverse and pure longitudinal modes of the vector boson contain the phase velocity tachyonic branch cuts. A complete calculation should include their effects undoubtedly. Compared in contrast with the Goldstone propagator, the imaginary part of the transverse and pure longitudinal branch cuts are not so concentrated around the $k^0 = \pm | \vec{k}|$ area. However, we can still apply the ``quasi-pole'' approximation to estimate their effects. Similar to (\ref{RDefinition}), we can define
\begin{eqnarray}
R_L (\gamma, \alpha) = \int_0^{1+\delta} \text{Im} \left[ \frac{x^2-1} {x^2-1+i \epsilon-\alpha + 2 \gamma (x^2-1)  (1-x Q_0(x))+i \epsilon} \right] dx, \nonumber \\
R_T(\gamma, \alpha) = \int_0^{1+\delta} \text{Im} \left[ \frac{1} {x^2-1+i \epsilon-\alpha -\gamma -  \gamma (x^2-1)  (1-x Q_0(x))+i \epsilon} \right] dx
\end{eqnarray}
to be the reduced ``residues'' of the ``quasi-poles'' of the longitudinal and transverse polarizations. Practical calculations show that $R_L(\gamma, \alpha) \ll R(\gamma, \alpha)$, while $R_T(\gamma, \alpha)$ is typically one or two orders of magnitude smaller than $R(\gamma, \alpha)$. Therefore in most of the cases, we can safely ignore them.

Photon and gluon are the only known massless vector bosons. Practical experiments on quark-gluon plasma could only generate the temperature of at most GeV scale. Our reliable knowledge on the cosmology does not go beyond the 1 MeV, which is the temperature scale of the big bang nucleosynthesis (For a review, see the corresponding chapters in Ref.~\cite{PDG}). Both of them are far below the mass threshold of a $W$/$Z$ boson, and the shift on $\gamma$-$Z$ mixing at this temperature is also negligible. That is probably the reason why a ``real'' or ``on-shell'' originally massive vector bosons in the thermal plasma have received so little attention in the literature, unlike the well-known dressed photon and gluons. However, Beyond the standard model (BSM) studies involve much higher temperature scales in the earlier universe.

For the dark matter freeze-out process, the typical temperature is usually far below the mass of the dark matter mass (See Ref.~\cite{WIMPReview} for a review). In fact, $T \sim \frac{m_{\text{DM}}}{26}$, where $m_{\text{DM}}$ is the dark matter mass. If, e.g.,  the dark matter annihilates into the massive vector bosons with their mass $m_V \ll m_{\text{DM}}$, these vector bosons can be regarded as the massless objects and the thermal corrections on masses do not affect the phase space integration significantly. If, on the other hand, $m_V \sim m_{\text{DM}}$, the freeze-out temperature is then too small compared with the $m_V$  for the significant thermal corrections, therefore they can still be neglected.

The feebly-interacting dark matter (FIMP) \cite{FreezeinAncester} is created in the higher temperature. The typical temperature for the freeze-in process is approximately of the same scale of the dark matter mass. Therefore, if the ``original mass'' of the participating massive vector boson is also in this scale, the thermal corrections on this vector boson might be non-ignorable.

Another possible application is the sterile neutrino production and decay in the early universe \cite{ZeroTempApprox, LightLeptogenesis}. Ref.~\cite{LaineNeutrino1, LaineNeutrino2, GarbrechtNeutrino} had calculated this in the unbroken phase. Although Ref.~\cite{LaineNeutrino3, BelowTcLaine2, BelowTcLaine3, BelowTcLaineNLO} performed the computation in the broken phase, they had only considered the mass range $\lesssim 20$ GeV in concert with their approximation $\mathcal{K}^2=0$. Detailed analysis confirmed that this is equivalent to getting rid of all the ``Goldstone boson fraction'' contributions, which is reliable in this mass range. If the mass of the sterile neutrino is comparable with the electro-weak phase transition temperature $\sim 100$ GeV, $W$/$Z$ gauge boson mass shifts and the Goldstone boson fractions might play crucial roles. This is important in the sterile neutrino portal dark matter models\cite{MyPaper1, MyPaper2, Hantao1, Hantao2, Spain1, Signals, LateExample}. Such a sterile neutrino can also induce the leptogenesis\cite{LightLeptogenesis}. In the previous literature, people use the zero-temperature mass of the vector bosons\cite{ZeroTempApprox}, or use some ansatz method\cite{LightLeptogenesis} to estimate these processes. Now, with the knowledge of the originally massive vector boson emotions in the thermal plasma, one can compute more precisely in the future.

\section{Summary}

In this paper, relied on a simple toy model and inspired by the convenient Goldstone equivalence gauge, we studied the behaviour of a massive vector boson in a thermal environment in the broken phase. The Goldstone equivalence gauge helps us decompose the transverse and longitudinal polarization contributions in the propagator, and after considering the self energy diagrams, we can examine in detail on how the longitudinal polarization vector spew out the Goldstone component. We also answered the question where the remained Goldstone boson goes in the thermal plasma shortly after the crossover. We find out that Goldstone boson degree of freedom was hidden in the tachyonic branch cut, and this branch cut can be treated as a quasi-pole to simplify the further calculations. We also show the external-leg Feynmann rules for the vector bosons as well as the (approximated, but physical) Goldstone boson. Gauge boson mixing case has also been discussed. It is then possible to calculate the tree-level processes involving the originally massive vector bosons with these Feynmann rules in the logic similar to the zero temperature situation.

\begin{acknowledgements}
We thank for Junmou Chen, Pyungwon Ko, Ligong Bian,  Fa-Peng Huang, Dong Bai, Chun Liu, Chen Zhang for helpful discussions. This work is supported by the Korea Research Fellowship Program through the National Research Foundation of Korea (NRF) funded by the Ministry of Science and ICT (2017H1D3A1A01014127).

\end{acknowledgements}

\appendix
\section{Discussions of the Extended Ward-Takahashi Identity in the Broken Phase} \label{WTIdentityAppendix}

Ward-Takahashi identity is the result of the gauge symmetry. This can be derived from the path integral method by applying infinitesmall changes on the field parameters in the integrands at the zero temperature. For the finite temperature situation, the only difference is the time parameter integration track, and other zero-temperature results are still available. Therefore, the gauge symmetry still leads to the Ward-Takahashi identity in the thermal plasma.

In the broken phase, there is still a version of Ward-Takahashi identity. Remember that the vev is also a part of the Higgs boson, and the gauge transformation operation requires the vev to transform as well, then Ward-Takahashi identity can also be derived from the path integral method. Rather than giving a complete proof\cite{WTIdentityBrokenPhase}, we only note that in the broken phase, the ``Noether current'' becomes
\begin{eqnarray}
j_{\mu} = j_{\mu}^{\text{vi}}+ i ( v \partial_{\mu} \phi - v^2 g A_{\mu}),
\end{eqnarray}
where $j_{\mu}^{\text{vi}}$ are the vev independent terms. A calculation of $\partial^{\mu} j_{\mu}$ gives $\partial_{\mu} \partial^{\mu} \phi$ and $\partial^{\mu} A_{\mu}$, and these can be replaced by the equations of motion. A direct calculation of the equations of motion shows that $\partial^2 \phi = \frac{\partial \mathcal{L}_{\phi \text{-coupling terms}}}{\partial \phi} + m_A \partial_{\mu} A^{\mu}$. Remember $m_A = g v$, so
\begin{eqnarray}
\langle \partial^{\mu} j_{\mu} \rangle = \langle \partial^{\mu} j_{\mu}^{\text{others}} \rangle + i v \times  \langle (\phi \text{-interaction terms}) \rangle.
\end{eqnarray}
Then we can follow the usual method to derive the Ward identity. Finally, $\partial^{\mu} \rightarrow -i k^{\mu}$, and $\phi$-interaction terms contribute to $m_A$ in the Goldstone component, so
\begin{eqnarray}
k_M \mathcal{M}^{M\dots} = 0,	\label{ModifiedWT}
\end{eqnarray}
where $k_M = (k_{\mu}, i m_A)$ for the inwards momentum.

Note that $m_A$ originates from the vev $g v$, therefore the $m_A$ in the $k_M$ does not depend on the 4-dimensional momentum values. This is important for deriving the (\ref{WTEquivalent}-\ref{CriticalEquation}), (\ref{MasslessGoldstoneRxi2}), (\ref{LandauLongitudinalPolarizationDerivation}).

\newpage
\bibliography{ThermalVector}
\end{document}